\documentclass[twocolumn]{aastex631}
\usepackage{color}
\usepackage[utf8]{inputenc}
\usepackage[T1]{fontenc}
\usepackage{amsmath}

\shortauthors{Zhu et al. 2025b}

\begin{document}
	
\title{Only Nitrogen-Enhanced Galaxies Have Detectable UV Nitrogen Emission Lines at High Redshift}

\author[0000-0002-1333-147X]{Peixin Zhu}
\affiliation{Center for Astrophysics $|$ Harvard \& Smithsonian, 60 Garden Street, Cambridge, MA 02138, USA}

\author[0000-0001-8152-3943]{Lisa J. Kewley}
\affiliation{Center for Astrophysics $|$ Harvard \& Smithsonian, 60 Garden Street, Cambridge, MA 02138, USA}
\affiliation{Research School of Astronomy and Astrophysics, Australian National University, Australia}

\author[0000-0003-4512-8705]{Tiger Yu-Yang Hsiao}
\affiliation{Center for Astrophysics $|$ Harvard \& Smithsonian, 60 Garden Street, Cambridge, MA 02138, USA}
\affiliation{Department of Astronomy, The University of Texas at Austin, 2515 Speedway, Austin, Texas 78712, USA}
\affiliation{Cosmic Frontier Center, The University of Texas at Austin, Austin, TX 78712}

\author[0000-0002-9081-2111]{James Trussler}
\affiliation{Center for Astrophysics $|$ Harvard \& Smithsonian, 60 Garden Street, Cambridge, MA 02138, USA}

\email{peixin.zhu@cfa.harvard.edu}

\begin{abstract}

The detections of bright UV nitrogen emission lines in some high-redshift galaxies suggest unexpectedly high nitrogen-to-oxygen ratios ($\log(\rm N/O)\gtrsim-1.0$) compared to local values ($\log(\rm N/O)\gtrsim-1.5$) at similar metallicities ($12+\log(\rm O/H)\lesssim8.0$). Although the presence of these `N-enhanced' galaxies indicates signatures of atypical chemical enrichment processes in the early universe, the prevalence of nitrogen enhancement in high-$z$ galaxies is unclear. So far, only $\sim$10 $z>5$ galaxies have nitrogen abundance measurements, and they all suggest elevated N/O ratios. Do all high-redshift galaxies exhibit elevated N/O ratios, or are we simply missing `N-normal' galaxies whose nitrogen abundances follow the local N/O scaling relation? To tackle these questions, we calculate the detection limits of UV \ion{N}{3}] or \ion{N}{4}] lines in current JWST surveys CEERS and JADES, and compare them to predictions from both `N-enhanced' and `N-normal' AGN narrow-line region and \ion{H}{2} region photoionization models. We find that CEERS can only detect galaxies with significant nitrogen enhancement ($\log(\rm N/O)\gtrsim-0.4$), while JADES can only detect galaxies with moderately elevated N/O ratios compared to local values ($\log(\rm N/O)\gtrsim-1.0$). Even for the deepest exposure in JADES, UV nitrogen lines produced by `N-normal' galaxies at $z>5$ are too faint and thus not detectable, making their nitrogen abundance unmeasurable. Our results suggest that the existing sample of galaxies with measurable nitrogen abundances at $z\gtrsim5$ is incomplete and biased toward galaxies with significantly elevated N/O ratios. Deep ($t_{\rm exp}\sim40-500\,$hs) spectroscopic surveys will be crucial for building a complete sample to study nitrogen enrichment mechanisms in the early universe.

\end{abstract}


\section{Introduction}

The presence of unexpectedly strong ultraviolet (UV) nitrogen emission lines \ion{N}{3}]$\lambda1747\_54$ and \ion{N}{4}]$\lambda\lambda1483,86$ in a small subset of high redshift galaxies \citep[e.g.][]{bunker_jades_2023,castellano_jwst_2024,napolitano_dual_2024} provides compelling evidence that the early universe was a chemically and physically distinct environment from the local universe. 
These {\color{black}`N-enhanced'} galaxies have elevated N/O ratios ($\log(\rm N/O)\gtrsim-1.0$) compared to the local values ($\log(\rm N/O)\gtrsim-1.5$, \citet{izotov_heavy-element_1999,2017MNRAS.466.4403N}) at similar metallicities ($12+\log(\rm O/H)\lesssim8.0$), posing new challenges to the established nitrogen enrichment mechanisms \citep{renzini_advanced_1981,van_zee_abundances_1998}. 

Proposed mechanisms for this atypical nitrogen enrichement include N-rich stellar winds from Wolf-Rayet stars \citep{watanabe_empress_2024}, massive stars ($M_{\star}\sim60-300\,M_{\odot}$) that experience dramatic mass-loss in the H-burning phase \citep{pascale_nitrogen-enriched_2023,topping_metal-poor_2024}, and supermassive stars ($M_{\star}\gtrsim1000\,M_{\odot}$) that formed in extremely dense clusters and collapse at the end of the H-burning \citep{charbonnel_n-enhancement_2023,marques-chaves_extreme_2024}. Gas flows could also reduce the oxygen abundance and lead to elevated N/O ratios through pristine gas accretion \citep{stiavelli_what_2025} or oxygen-rich galactic winds \citep{rizzuti_high_2025}. Some of these high-redshift `N-enhanced' galaxies are also interpreted as proto-clusters that later evolve into globular clusters in the local universe \citep{cameron_nitrogen_2023,senchyna_gn-z11_2023,ji_connecting_2025}.

In contrast to the extensive discussion on nitrogen enrichment mechanisms, the prevalence of nitrogen enhancement in the early universe remains underexplored. At z$\gtrsim7$, UV nitrogen emission lines are the only probes for nitrogen abundance measurement because optical nitrogen emission lines are redshifted beyond the observed wavelength range of JWST NIRSpec. However, among over 500 confirmed $z>7$ galaxies, less than 10 exhibit detectable UV nitrogen emission lines and all indicate significantly or moderately elevated N/O ratios ($\log(\rm N/O)\gtrsim-1.0$) \citep{ji_connecting_2025}. The absence of UV nitrogen emission lines in the observed spectra of all remaining high-redshift galaxies raises the question of whether these remaining galaxies have elevated nitrogen abundance or not. 

To tackle this question, we use the `N-enhanced' and `N-normal' AGN narrow-line region (NLR) and \ion{H}{2} region photoionization models developed by Zhu et al. (2025, submitted) to estimate the minimum nitrogen abundance required for UV nitrogen lines (such as \ion{N}{3}] and \ion{N}{4}]) to be detectable in the JWST observations. {\color{black}We include AGN NLR models because several high-$z$ “N-enhanced” galaxies show AGN signatures \citep[e.g.,][]{ji_ga-nifs_2024,napolitano_dual_2024,curti_jades_2025}, and the harder AGN ionizing spectra can alter the detectability of UV nitrogen lines and thus the inferred abundance thresholds.} These models allow us to place constraints on the nitrogen abundances of high-redshift galaxies that lack detected UV nitrogen lines, offering insight into the completeness and potential bias of the current N/O ratio census in the early universe.

This letter is structured as follows: Section~\ref{sec:data} describes the observational data, and Section~\ref{sec:model} provides a brief overview of the photoionization models. Section~\ref{sec:detection} presents the estimates of detection limits. Results are shown in Section~\ref{sec:result}, followed by a discussion in Section~\ref{sec:discussion}.

\section{Observational Data}\label{sec:data}

To provide an independent test of our estimated detection limits in NIRSpec PRISM, we compile a `JWST archival' sample from the DAWN JWST Archive (DJA; {\color{black}see \citet{graaff_rubies_2025,heintz_jwst-primal_2025}}), where all public JWST/NIRSpec spectra have been uniformly reduced using the msaexp pipeline. {\color{black}From the DJA NIRSpec Merged Table v4.4 \citep{brammer_2025_15472354,valentino_gas_2025},} 
we identify 3917 galaxies in the redshift range $4.5<z<15$ with available PRISM spectra and high-quality spectral fits (grade = 3).

To present the detection limits of current JWST observational data, we select galaxies whose \ion{N}{3}]$\lambda1747\_54$ or \ion{N}{4}]$\lambda\lambda1483,86$ emission lines have signal-to-noise ratios greater than 3 (S/N$>$3) and positive equivalent width (EW). We compare the observations with photoionization models using EW-based UV diagnostic diagrams, as these diagrams provide the most effective separation between AGN- and \ion{H}{2}-dominated spectra (Zhu et al. 2025, submitted). To place these galaxies on the EW(\ion{N}{3}])$-$\ion{N}{3}]/\ion{He}{2} and EW(\ion{N}{4}])$-$\ion{N}{4}]/\ion{He}{2} diagnostic diagrams in Section~\ref{sec:result}, we further require a detection of \ion{He}{2}$\lambda1640$ at S/N$>$2. We note that this additional S/N cut on \ion{He}{2} does not affect the flux or EW distribution of the selected nitrogen emission lines. These criteria yield 8 galaxies with \ion{N}{3}] detections and 5 galaxies with \ion{N}{4}] detections. {\color{black}Cross-matching our selection with literature samples \citep[e.g.,][]{ji_connecting_2025} identifies a single galaxy in common: GN-z11.} For all selected galaxies, we convert their observed EWs into rest-frame EWs following $\rm EW_{rest}=EW_{obs}/(1+z)$ for the comparisons in Section 5. 

Similarly, we applied the same S/N$>$3 and EW$>$0 criteria to the `JWST archival' sample for \ion{C}{3}]$\lambda\lambda1907,9$, \ion{C}{4}$\lambda\lambda1549,51$, \ion{He}{2}$\lambda1640$, and \ion{O}{3}]$\lambda\lambda1661,6$ emission lines to obtain subsamples with $3\sigma$ detection in each of these lines individually. We also require S/N$>$2 \ion{He}{2}$\lambda1640$ detections for all selected galaxies because \ion{He}{2} are involved in all EW-based UV diagnostic diagrams, and this criterion does not affect the flux or EW ranges of other emission lines.

To provide a validation for our `N-enhanced' photoionization models, we include the $z>5$ `N-enhanced' galaxy sample compiled in Zhu et al. (2025, submitted), which consists of eight $z>5$ `N-enhanced' galaxies reported to exhibit significantly elevated N/O ratios ($\log(\rm N/O)\gtrsim-1.0$) and have $\geq3\sigma$ UV \ion{N}{4}] detections. Due to the limited \ion{He}{2} detections in this sample, only four galaxies are included in this work: GHZ2 \citep{castellano_jwst_2024}, GHZ9 \citep{napolitano_dual_2024}, GS-z9-0 \citep{curti_jades_2025}, and RXCJ2248-ID \citep{topping_metal-poor_2024}.

UV observations from the Cosmic Origins Spectrograph (COS) Legacy Spectroscopic Survey (CLASSY, \citet{berg_cos_2022}) are also included to provide an independent test for our `N-normal' \ion{H}{2} region models. The CLASSY sample includes 45 nearby ($0.002<z<0.182$) UV-bright star-forming galaxies whose UV emission line fluxes and EWs have been carefully measured in \citet{mingozzi_classy_2022}. We include CLASSY galaxies in our UV diagnostic diagrams whenever the corresponding measurements are available and have S/N$>$3.

\section{Photoionization Models}\label{sec:model}

We utilize the photoionization models developed in Zhu et al. (2025, submitted) to predict the emission line fluxes and EWs of `N-enhanced' and `N-normal' galaxies. We consider both \ion{H}{2} region and AGN NLR photoionization models to account for the changes in emission line brightness caused by the change of ionizing sources. {\color{black}We do not include the shock model because, by itself, it cannot consistently reproduce the full set of observed spectral features (e.g., line ratios and EWs) of `N-enhanced' galaxies (Zhu et al., in prep). However, a partial contribution from shocks, if present, could generate brighter UV nitrogen lines at fixed N/O ratio relative to the \ion{H}{2} and AGN models \citep{flury_new_2024}.}

For the \ion{H}{2} region photoionization model, the stellar ionizing spectra are generated by \citet{li_massmetallicity_2024} using the Stromlo Stellar Tracks \citep{grasha_stromlo_2021} stellar evolution models with rotation velocity $v/v_{\rm crit}=0.4$ and computed with the Flexible Stellar Population Synthesis (FSPS, \citet{conroy_propagation_2009,conroy_propagation_2010}). These stellar ionizing spectra assume a Chabrier initial mass function \citep{chabrier_galactic_2003} with a mass range of $10\leq \rm M_{\star}/M_{\odot}\leq300$, and {\color{black}a continuous star formation history with a constant star formation rate}. The CMFGEN Wolf-Rayet (WR) Stars spectra from \citet{smith_realistic_2002} are included when WR stars are identified (stars have $T_{\rm eff}>10^4\,$K and the surface hydrogen mass fraction $X_{\rm surf}<0.3$) in FSPS. The stellar ionizing spectra are taken from the 5 Myr snapshots.

For the AGN NLR photoionization model, the AGN ionizing spectra are generated using OXAF \citep{thomas_physically_2016} by varying the peak energy of the accretion disk thermal emission ($E_{\rm peak}$) in the range of $-2.0\leq\log (E_{\text{peak}}/\text{keV})\leq-1.0$. These AGN ionizing spectra are valid for black holes (BHs) that are undergoing the thin-disk accretion mode, which typically have black hole masses of $10^6\lesssim \rm M_{\rm BH}/M_{\odot}\lesssim10^9$ and accretion rates of $0.05\lesssim L/L_{\rm Edd} \lesssim0.3$ \citep {novikov_astrophysics_1973,done_intrinsic_2012,thomas_physically_2016}. {\color{black}These black hole mass and accretion rate ranges are consistent with those inferred for JWST-discovered high-$z$ AGNs \citep[e.g.,][]{harikane_jwstnirspec_2023,matthee_little_2024,juodzbalis_jades_2025}.}

All photoionization models are calculated with MAPPINGS version 5.2 \citep{binette_radiative_1985,sutherland_cooling_1993,dopita_new_2013,sutherland_mappings_2018}, assuming an isobaric (constant-pressure) structure and including dust depletion effects using the factors from \citet{jenkins_unified_2009}. {\color{black}We note that adopting dust-free models or depletion patterns from the Magellanic Clouds \citep{roman-duval_metal_2022} will not change our results. Because dust-free and dusty models have very similar distribution at low metallicity (12+$\log(\rm O/H)\leq8.4$) on the UV diagnostic diagrams (Zhu et al.2025, submitted) and optical diagnostic diagrams \citep{zhu_new_2023}.} Abundance scaling relations in \citet{2017MNRAS.466.4403N} are used in `N-normal' models to predict the emission line fluxes and EWs for galaxies that follow the local nitrogen abundance scaling relation. The `N-enhanced' models are generated by simply increasing the nitrogen abundance in `N-normal' models to the range of $-0.8\lesssim\log(\rm N/O)\lesssim0.4$ with a $0.1\,$dex interval. All photoionization models have a varied range of gas-phase metallicity ($7.2\leq$12+$\log(\rm O/H)\leq8.7$), ionization parameter ($-4.0\leq\log(U)\leq-0.5$), and gas pressure (5.0$\lesssim\log{(P/k)}\lesssim$9.8) to accommodate a wide range of physical environments in various galaxies. {\color{black}The gas pressure range corresponds to electron densities ($n_e$) of $10^1\lesssim n_e\,(\rm cm^{-3})\lesssim10^6$ for $T_e\approx10^4\,$K. The high-pressure limit is chosen to encompass the extreme $n_e$ reported in some high-$z$ `N-enhanced' galaxies \citep[e.g.][]{larson_ceers_2023,senchyna_gn-z11_2023,maiolino_small_2024,topping_metal-poor_2024}.}

\section{Detection Limit Estimates for JWST observations}\label{sec:detection}

We estimate the detection limit of JWST observation on the UV emission lines using the limiting sensitivity curve\footnote{Figure~2 in https://jwst-docs.stsci.edu/jwst-near-infrared-spectrograph/nirspec-performance/nirspec-sensitivity\#gsc.tab=0} of the NIRSpec PRISM spectra in the multi-object spectroscopy (MOS) mode presented in the JWST User Document \citep{2016jdox.rept......}. {\color{black}We use PRISM rather than the gratings because its limiting continuum sensitivity is about an order of magnitude deeper. The impact of using the greater line sensitivity of the gratings is discussed in Section~\ref{sec:discussion}.}

At $5\lesssim z\lesssim 13$, UV nitrogen lines \ion{N}{3}] and \ion{N}{4}] are redshifted to the observed wavelength range of $\lambda_{\rm obs}=1-2.5\,\rm \mu m$, where the PRISM sensitivity curve has the minimal limiting sensitivity value ($f_{\nu}\approx10^{-4}\,$milli-Jy) and the best detectability. The limiting sensitivity value represents the flux that achieves S/N = 10 in 10$\,$ks exposure. To compute the noise level of a 10$\,$ks exposure, we first convert the $f_{\nu}$ into $f_{\lambda}$ via:
\begin{equation*}
f_{\lambda} = f_{\nu}\times c/{\lambda_{\rm obs}^2} = 3\times 10^{-20}/\lambda_{\rm obs}^2\,\rm (erg\,s^{-1}\,cm^{-2} \text{\AA}^{-1})
\end{equation*}
where the $\lambda_{\rm obs}$ is in units of $\rm \mu m$ throughout this calculation. The noise level of a 10 ks exposure using PRISM in MOS mode is then $\sigma = f_{\lambda}/10$. 

For an observation that has an exposure time of $t'\,$ks, the corresponding noise level is $\sigma'=\sigma \times \sqrt{10/t'}$. The $3\sigma$ flux of an emission line is then:
\begin{equation*}
	\rm Flux_{3\sigma} = 3\times\sigma'\times\Delta\lambda 
\end{equation*} 
The $\Delta\lambda$ is the observed line width of the emission line, with the minimum line width determined by the spectral resolution R through:
\begin{equation*}
\Delta\lambda\,(\text{\AA})\gtrsim\lambda_{\rm obs}\,(\mu m)\times10^4/R
\end{equation*}

For an emission line to be detected at $>3\sigma$ level in a $t'\,$ks exposure observed by JWST NIRSpec PRISM spectrum, its minimum flux should be:
\begin{equation*}
	\text{Flux}_{3\sigma,\rm min} = 9.0\times10^{-17}\times \sqrt{10/t'}/\lambda_{\rm obs}/R\,\rm (erg\,s^{-1}\,cm^{-2})
\end{equation*} 

{\color{black}We adopt the spectral resolution R for a fully illuminated aperture from \citet{2016jdox.rept......}, where R$\approx30-70$ for $\lambda_{\rm obs}=1-2.5\,\rm \mu m$ in PRISM. However, \citep{de_graaff_ionised_2024} notes that for point source (e.g., compact high-$z$ galaxies), the actual spectral resolution is higher by nearly a factor of 2.}

We then calculate the $3\sigma$ flux of the faintest UV emission lines that can be detected in a $t'\,$ks exposure. We estimate the minimum detected $3\sigma$ flux for PRISM spectra in two JWST surveys: The Cosmic Evolution Early Release Science Survey (CEERS, \citet{finkelstein_cosmic_2025}) and The JWST Advanced Deep Extragalactic Survey (JADES, \citet{eisenstein_overview_2023}), to obtain representative estimates for the detection limits in most publicly available JWST data.  The typical exposure time for CEERS NIRSpec PRISM observations is 3107.4 s, while the maximum exposure time in JADES Deep observations is {\color{black}27.7 hours \citep{scholtz_jades_2025}.} We also convert the estimated $3\sigma$ flux into EW, adopting the average rest-frame continuum level ($\sim10^{-19}~\mathrm{erg\,s^{-1}\,cm^{-2}\,\text{\AA}^{-1}}$) of the $z>5$ `N-enhanced' galaxy sample compiled in Zhu et al. (2025, submitted). The resulting $3\sigma$ flux and rest-frame EW detection limits are listed in Table~\ref{tab:1}. 

We note that all $z>5$ `N-enhanced' galaxies are relatively bright compared to regular galaxies at the high-$z$ redshift. Regular galaxies with fainter continuum levels would have higher $3\sigma$ flux and rest-frame EW detection limits than the estimation in Table~\ref{tab:1}.

\setlength{\tabcolsep}{3.0pt}
\begin{deluxetable}{c|c|c|c|c}[hbt]
\centering
\tablewidth{1pc}
\tablecaption{Detection Limits of UV Emission Lines with PRISM in the CEERS and JADES Surveys. \label{tab:1}}
\tablenum{1}
\tablehead{
\colhead{} &  
\multicolumn{2}{c|}{\textbf{CEERS}} & 
\multicolumn{2}{c}{\textbf{JADES (Deep)}} }
\startdata
$t_{\rm exp}$ & \multicolumn{2}{c|}{3.1 ks} & \multicolumn{2}{c}{99.7 ks} \\
\hline
$\lambda_{\rm obs}$ & 1 $\mu$m & 2.5 $\mu$m & 1 $\mu$m & 2.5 $\mu$m \\
$R$ & 30 & 70 & 30 & 70 \\
\hline
Flux$_{3\sigma,\rm min}$\tablenotemark{\scriptsize a} & 5.3$\times10^{-18}$ & 9$\times10^{-19}$ & {\color{black}9$\times10^{-19}$} & {\color{black}2$\times10^{-19}$} \\
EW$_{3\sigma,\rm min}$\tablenotemark{\scriptsize b}
	& $\sim53$ \AA &  $\sim9$ \AA &  $\sim9$ \AA &  $\sim2$ \AA \\
\enddata
\tablenotetext{\scriptsize a}{Unit: $\mathrm{erg\,s^{-1}\,cm^{-2}}$}
\tablenotetext{\scriptsize b}{At rest-frame continuum level of $\sim10^{-19}\,$erg$\,\rm s^{-1}\,cm^{-2}\,\text{\AA}^{-1}$.}
\end{deluxetable}
\vspace{-1.em} 

\section{Results: Observational Bias on UV Emission Lines}\label{sec:result}

\begin{figure*}[hbt]
\epsscale{1.2}
\plotone{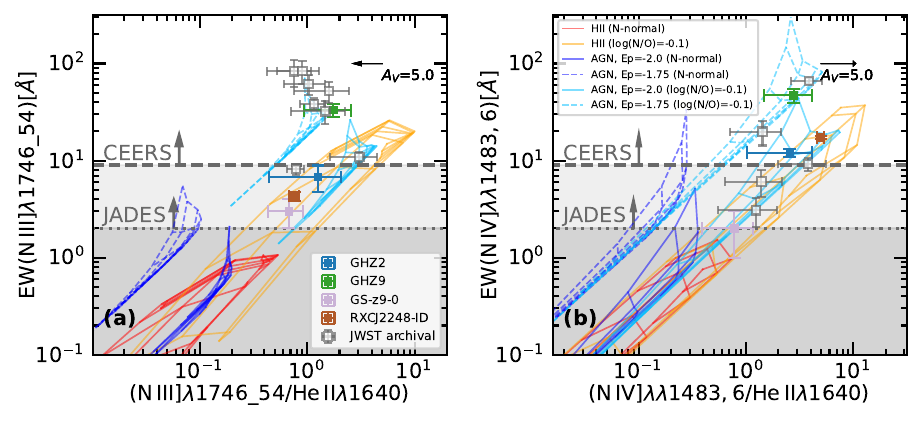}
\caption{Detection limits of JWST observations for UV nitrogen emission lines \ion{N}{3}] and \ion{N}{4}]. The gray dashed and dotted lines indicate the 3$\sigma$ detection thresholds for the CEERS survey (exposure time: 3.1 ks) and the JADES Deep survey (maximum exposure time: 27.7 hours), respectively, assuming a rest-frame continuum level of $\sim10^{-19}\,\rm erg/cm^2/s/${\AA}. \ion{H}{2} region models at $\log(P/k) = 9.2$ are shown in red (N-normal) and orange (N-enhanced), while AGN photoionization models at the same gas pressure are shown in blue (N-normal) and deepskyblue (N-enhanced). Solid and dashed lines represent AGN models with $\log(E_{\text{peak}}/\rm{keV}) = -2.0$ and $-1.75$, respectively. Each model grid consists of constant metallicity lines ($12 + \log(\mathrm{O/H}) = 7.30, 7.70, 8.00, 8.28, 8.43$ for AGN; $7.15, 7.75, 8.15, 8.43$ for \ion{H}{2}) and constant ionization parameter lines (spanning $\log(U)=-3.0$ to $-0.5$ for AGN and $\log(U)=-4.0$ to $-0.5$ for \ion{H}{2}, in steps of 0.5 dex). Large colored squares represent the $z > 5$ `N-enhanced' galaxy sample compiled in Zhu et al. (2025, submitted), and open gray squares represent archival $>3\sigma$ \ion{N}{3}] and \ion{N}{4}] detections collected from the DJA. Black arrows in the upper-right corner of each panel indicate the effect of $A_V = 5.0$ mag dust extinction on the UV observations derived from the extinction curve in \citet{cardelli_relationship_1989}.
\label{fig:1}}
\end{figure*} 

UV nitrogen emission lines are subject to significant observational bias due to the detection limit of current JWST observations. We demonstrate this observational bias by comparing the calculated detection limits on UV emission line equivalent widths in Table~\ref{tab:1} with photoionization model predictions on the EW(\ion{N}{3}])-\ion{N}{3}]/\ion{He}{2} and EW(\ion{N}{4}])-\ion{N}{4}]/\ion{He}{2} diagnostic diagrams. 

We present the minimum EW$_{3\sigma,\rm min}$ detection limit of the CEERS survey and JADES Deep observations, shown as dashed and dotted lines in Figure~\ref{fig:1}. These limits are obtained at $\lambda_{\rm obs}=2.5\,\mu$m, corresponding to galaxies at $z\approx13$ for \ion{N}{3}] and $z\approx16$ for \ion{N}{4}]. For galaxies at $z=10$, the EW$_{3\sigma,\rm min}$ thresholds for \ion{N}{3}] and \ion{N}{4}] emission lines increase by factors of $\sim1.8$ and $\sim2.7$, respectively.

As shown in Figure~\ref{fig:1}, photoionization models with normal (local) nitrogen abundance (red and blue grids) fall far below the detection limit of the CEERS survey, and the majority are also below the detection limit of JADES Deep observations. This result indicates that the CEERS survey can only detect UV nitrogen lines in galaxies that have significantly elevated N/O ratios ($\log(\rm N/O)\gtrsim-0.4$), and JADES Deep observations can reveal UV nitrogen lines in galaxies with moderately enhanced nitrogen abundance ($\log(\rm N/O)\gtrsim-1.0$). {\color{black}Indeed, our estimated CEERS detection limit is consistent with \citet{isobe_jwst_2023}, who converted 3$\sigma$ upper limits of the \ion{N}{3}] flux into N/O upper limits ($\log(\rm N/O)_{UL}$) and found $\log(\rm N/O)_{UL}\gtrsim-0.4$ for 50 CEERS galaxies without \ion{N}{3}] detections.}

In contrast, star-forming galaxies and galaxies hosting weak AGNs ($\log (E_{\text{peak}}/\text{keV})\approx-2.0$) with normal nitrogen abundance are unable to produce sufficiently bright UV nitrogen emission lines to be detected in either the CEERS and JADES surveys.

\begin{figure*}[hbt]
\epsscale{1.2}
\plotone{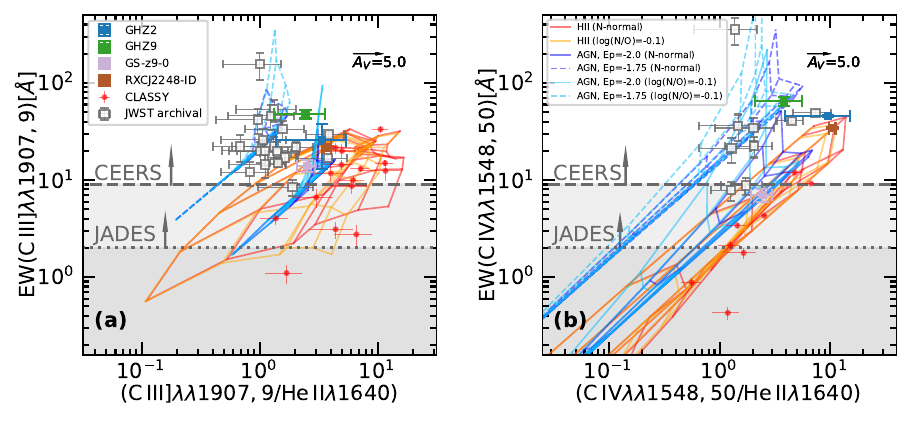}
\caption{Detection Limits of JWST observations for the UV emission lines \ion{C}{3}] and \ion{C}{4}. CLASSY galaxies with \ion{C}{3}] and \ion{C}{4} detections are included for comparison and shown as red circles. Model grid parameter ranges and symbols follow those in Figure~\ref{fig:1}. 
\label{fig:2}}
\end{figure*} 

The only exception for a galaxy with normal nitrogen abundance to exhibit sufficiently bright UV nitrogen lines reaching the detection limit of JADES Deep observations is when it hosts a strong AGN. As shown in Figure~\ref{fig:1}, `N-normal' AGN models with $\log (E_{\text{peak}}/\text{keV})\geq-1.75$ exceed the JADES detection threshold at $\log(U)\gtrsim-1.5$ and $\log(P/k)\gtrsim9.2$. These AGN models correspond to galaxies hosting a central black hole with $\log(M_{\rm BH}/M_{\odot})=8.0$, accreting at an Eddington ratio of $f_{\rm Edd}\geq0.3$, and embedded in a dense ($n_e\gtrsim10^5\,\rm cm^{-3}$) and compact environment, where the electron desntiy $n_e$ is converted from gas pressure $P$ using $P=nkT$. 

The JWST observations in Figure~\ref{fig:1} also support our detection limit estimations for UV nitrogen lines. Both the `N-enhanced' galaxy sample compiled by Zhu et al. (2025, submitted) and the JWST archival $S/N>3$ observations lie above the detection threshold of JADES Deep observations. The deepest existing observation on \ion{N}{4}] to date is for GS-z9-0, which was observed with PRISM for $\sim252$ ks \citep{curti_jades_2025}. This exposure time is more than twice the maximum exposure time of the JADES Deep survey, pushing the detection threshold further down to EW$_{3\sigma,\rm min}\sim1.2\,$\AA, which is just enough to reveal the \ion{N}{4}] emission lines in GS-z9-0. 

On the contrary, UV carbon emission lines \ion{C}{3}]$\lambda\lambda1907,9$ and \ion{C}{4}$\lambda\lambda1549,51$ are less affected by the detection limit of current JWST observations. As shown in Figure~\ref{fig:2}, both AGN and \ion{H}{2} region photoionization models adopting local carbon abundance are capable of producing \ion{C}{3}] and \ion{C}{4} lines that are bright enough to exceed the detection thresholds of CEERS and JADES surveys. In CEERS observations, galaxies with $\log(U)\gtrsim-3.0$ can produce detectable \ion{C}{3}] emission, while those with $\log(U)\gtrsim-1.5$ can produce detectable \ion{C}{4} emission. In the deeper JADES observations, the detectable range extends further: \ion{C}{3}] and \ion{C}{4} become visible in galaxies with $\log(U)\gtrsim-4.0$ and $\log(U)\gtrsim-2.5$, respectively.

\begin{figure*}[hbt]
\epsscale{1.2}
\plotone{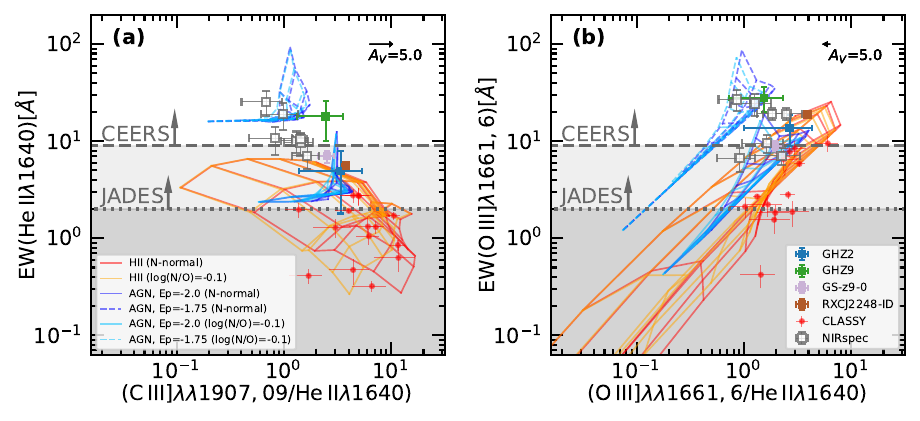}
\caption{Detection Limits of JWST observations for the UV emission lines \ion{He}{2} and \ion{O}{3}]. CLASSY galaxies with $>3\sigma$ \ion{O}{3}] and \ion{He}{2} detections are included for comparison and shown as red circles. Model grid parameter ranges and symbols follow those in Figure~\ref{fig:1}. 
\label{fig:3}}
\end{figure*} 

Similar to nitrogen lines, \ion{He}{2}$\lambda1640$ also suffers from observational bias induced by the detection limit of current JWST observations. As shown in Figure~\ref{fig:3}(a), only AGN models with $\log(E_{\text{peak}}/\text{keV})\geq-1.75$ can produce sufficiently strong \ion{He}{2} emission to be detected in the CEERS survey. The JADES Deep survey can go deeper and reveal the \ion{He}{2} emission from weaker AGNs or star-forming galaxies that have extremely low metallicity ($12+\log(\rm O/H)\lesssim7.8$). 

\ion{O}{3}]$\lambda\lambda1661,6$ suffers less observational bias caused by these detection limits, with both AGN and \ion{H}{2} region photoionization models exceeding the detection limits of the CEERS and JADES surveys in Figure~\ref{fig:3}(b). For galaxies with $\log(U)\gtrsim-1.5$ and $\log(P/k)\gtrsim9.2$, their UV \ion{O}{3}] emission lines are detectable in most JWST observations with similar depth as the CEERS survey. For galaxies with less extreme environments (lower ionization parameter $-1.5\gtrsim\log(U)\gtrsim-2.5$ or lower gas pressure $9.2\gtrsim\log(P/k)\gtrsim7.0$), their \ion{O}{3}] emission lines can still be detected by JADES Deep observations. 

\section{Discussion}\label{sec:discussion}

By combining JWST sensitivity limits and our latest photoionization models for AGN and star-forming galaxies, we have shown that the UV nitrogen emissions from galaxies with normal (local) nitrogen abundance are too weak to exceed the detection limit of current JWST observations. Only `N-enhanced' ($\log(\rm N/O)\gtrsim-1.0$) galaxies emit strong enough UV nitrogen emission lines that are detectable in current JWST observing depth. Therefore, it is not surprising that among the $\gtrsim2000$ confirmed $z>5$ galaxies, only $\sim10$ galaxies have $>3\sigma$ detections on UV \ion{N}{3}] and \ion{N}{4}] emission lines, and they are all `N-enhanced' galaxies ($\log(\rm N/O)\gtrsim-1.0$). 

Current JWST observations indicate an occurence rate of $\lesssim5\%$ for `N-enhanced' phenomena ($\log(\rm N/O)\gtrsim-1.0$) in $z>5$ galaxies. The remaining $\sim95\%$ high-redshift galaxies have N/O ratios that are either closer to or equal to the local values ($\log(\rm N/O)\approx-1.5$). {\color{black}Some stacked spectra of $z>4$ galaxies have successfully revealed faint nitrogen emission lines that are consistent with mildly ($\log(\rm N/O)\approx-1.0$) to moderately ($\log(\rm N/O)\approx-0.4$) elevated N/O ratios \citep{hayes_average_2025,isobe_jades_2025}.} However, for galaxies with normal (local) nitrogen abundances, the existing stacked spectra are not yet deep enough to detect their nitrogen emission lines.

\begin{figure*}[hbt]
\epsscale{1.2}
\plotone{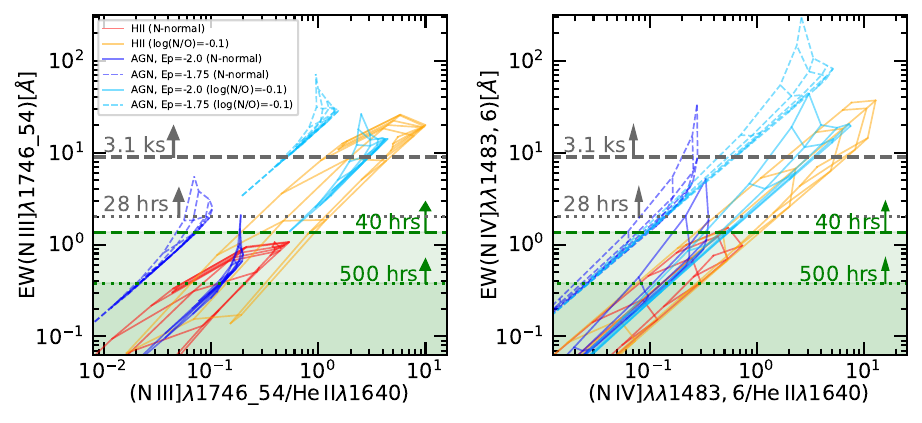}
\caption{Detection limit on \ion{N}{3}] and \ion{N}{4}] emission lines for JWST observations with different exposure times. The green dashed and dotted lines represent the 3$\sigma$ detection limit of a 40-hour survey and a 500-hour survey using NIRSpec PRISM, respectively, for galaxies with rest-frame continuum level at $\sim10^{-19}\,\rm erg/cm^2/s/${\AA}. Model grid parameter ranges follow those in Figure~\ref{fig:1}.
\label{fig:4}}
\end{figure*} 

Nevertheless, uncovering the mechanisms driving high N/O ratios in the early universe is essential. To uncover a more complete N/O distribution at $z>5$, a deeper spectroscopic survey using PRISM with exposure time exceeding 40 hours is required (Figure~\ref{fig:4}). {\color{black}To capture a larger sample of `N-normal' galaxies at high redshift, including star-forming with high gas pressure ($\log(P/k)\gtrsim7.0$) and high $\log(U)\gtrsim-1.5$) in addition to AGN-host galaxies}, deeper observations with $t_{\rm exp}\approx500\,$hours are required. 

{\color{black}Indeed, the stacked PRISM spectra of $z=4-10$ galaxies in \citet{hayes_average_2025} illustrate the need for long exposure time. Three stacks with total exposure times of $t\approx700\,$hours reveal UV \ion{N}{4}] and \ion{N}{3}] in galaxies with EW([\ion{O}{3}])$>500\,$\AA. Their inferred N/O values are $\log(\rm N/O)\approx-1.0$ for stacks with EW([\ion{O}{3}])$>1000\,$\AA, and $\log(\rm N/O)\approx-0.4$ for the stack with EW([\ion{O}{3}])$=500-1000\,$\AA. Galaxies with larger EW([\ion{O}{3}]) tend to have harder ionizing spectra (e.g., from recent starbursts or AGNs), which boosts UV nitrogen line strengths at fixed N/O. As a result, galaxies with lower N/O can still be detected in the higher EW([\ion{O}{3}]) stacks, resulting in a lower stack-average N/O estimate. On the contrary, the other two stacks with EW([\ion{O}{3}])$<500\,$\AA, despite twice the exposures time ($t\approx1400\,$hours), show no UV nitrogen line detections. These stacks have softer ionizing fields and lower ionization parameters ($\log(U)\lesssim-2.5$), producing very faint UV nitrogen lines (EW(\ion{N}{3}]$<0.1$) that remain below the detection limit even at $t_{\rm exp}=1400\,$hours.}

To more efficiently detect UV nitrogen lines in bright `N-normal' galaxies at $z>5$, a viable alternative is to use the $R\sim1000$ gratings G140M and G235M instead of PRISM at $\lambda_{\rm obs}=1-2.5\,\mu$m. Although the gratings have limiting continuum sensitivities that are about an order of magnitude higher than PRISM, their substantially greater spectral resolution ($R\sim1000$ compared to $R=30$–$70$) provides a factor of $\sim2-$2.5 improvement in line sensitivity \citep[also see Figure 9 in][]{eisenstein_overview_2023}. This increased line sensitivity translates into a reduction in the required exposure time by a factor of $\sim4-$6.25.

Another observational bias introduced by the current detection limit concerns the dominant excitation mechanism in identified `N-enhanced' galaxies. At a given nitrogen abundance, galaxies hosting strong AGN ($\log(E_{\text{peak}}/\text{keV})\geq-1.75$, {\color{black}corresponding to black holes with $\rm M_{\rm BH}/M_{\odot}\lesssim10^8$ accreting at $\log(L/L_{\rm Edd})\gtrsim-2.0$, consistent with most high-$z$ AGNs in \citet{maiolino_jwst_2025}}) produce brighter UV nitrogen emission lines and are therefore more likely to be detected in JWST observations. It is thus unsurprising that the current high-redshift `N-enhanced' galaxy sample includes more AGN-host galaxies than pure star-forming galaxies. Indeed, our recent study in Zhu et al. (2025, submitted) finds that 7 out of 8 known `N-enhanced' galaxies at $z>5$ are AGN-dominated. Similarly, \citet{isobe_jades_2025} study the stacked spectrum of 20 high-redshift Type-1 AGNs and find a $2.6\sigma$ \ion{N}{3}] line detection in the stacked spectrum. As a comparison, they find no \ion{N}{3}] detection in the stacked spectrum of $\sim600$ star-forming galaxies in the same redshift range.

The prevalence of AGN-dominated `N-enhanced' galaxies has led to speculations that nitrogen enrichment might be physically linked to AGN activity \citep{maiolino_small_2024,isobe_jades_2025}. Constructing an unbiased sample of `N-enhanced' galaxies at $z>5$ will be essential to determine whether the association between AGN and nitrogen enhancement is intrinsic or simply a consequence of detection bias in current JWST observations.

\section{Acknowledgement}
We thank the anonymous referee for thoughtful and useful comments, which help improved this paper.
JAAT acknowledges support from the Simons Foundation and \emph{JWST} program 3215. Support for program 3215 was provided by NASA through a grant from the Space Telescope Science Institute, which is operated by the Association of Universities for Research in Astronomy, Inc., under NASA contract NAS 5-03127.

This work utilizes data retrieved from the Dawn JWST Archive (DJA). DJA is an initiative of the Cosmic Dawn Center (DAWN), which is funded by the Danish National Research Foundation under grant DNRF140.

\bibliography{Paper_UV_Lim.bib}{}
\bibliographystyle{aasjournal}

\end{document}